# Structured Analysis Reveals Fundamental Mathematical Relationships between Wind and Solar Generations and the United Kingdom Electricity System

Anthony D Stephens[1] and David R Walwyn[2]
(correspondence to tonystephensgigg@gmail.com)

**Abstract**

The United Kingdom (UK) electricity system is recorded every five minutes, its records being available on the Gridwatch™ website. The paper explains how a set of complementary models, whose aims are to predict the behaviour of future electricity systems, may be based on suitably scaled historic records. The model predictions are first demonstrated for a scenario for the year 2035, proposed by the National Grid in FES 2022. An annual histogram of wind plus solar generation for this 2035 scenario reveals that a large amount of available wind and solar generation will exceed the ability of the electricity system to accommodate it. The dynamic models show that this excess generation will be far too variable to be of beneficial use, resulting in a reduction in the efficiency with which wind and solar generations are able to decarbonise the electricity system. It is shown how the dynamic models may be combined to produce a Compound Model which predicts the annual performance of future electricity systems. These predictions include an efficiency measure, the Marginal Decarbonisation Efficiency, (MDE), which quantifies the efficiency with which wind generation is able to decarbonise the electricity system, and is important because it is likely to determine the upper economic deployment of wind generation. It is explained how a carefully designed study of the behaviour of the electricity system using the Compound Model led to the formulation of a Generic Model. This provides a means of defining, in simple mathematical terms, the relationships between future electricity systems and wind and solar generations. Since MDE is an input variable to the Generic Model, the model enables predictions to be made of the optimal amount of wind generation needed to satisfy future electricity systems, and of the minimum amount of dispatchable generation required to compensate for the inherent variability of wind and solar generations. Dispatchable generation which causes the minimum amount of carbon dioxide emissions is likely to be provided by combined cycle gas turbines, but this is at variants with the predictions of the National Grid's steady state model, which see no need for gas generation in 2035. The National Grid's view is not supported by Germany and California, who do see an on-going need for gas generation to ensure security of their electricity systems during periods when there is insufficient wind and solar generation to do so.

**Nomenclature**

| | |
|---|---|
| **GW** | measure of generation in gigawatts |
| **GWw** | wind generation in GW accommodated by the electricity system |
| **GWw+s** | wind plus solar generation in GW accommodated by the electricity system |
| **Hdrm** | the portion of electrical demand available for wind/solar generation to satisfy |
| **MDE** | marginal decarbonisation efficiency, which is the reduction in carbon dioxide emissions, in million tonnes (MT)/annum, resulting from an additional GW of wind generation available from the wind fleet |
| **sm** | multiple of solar generation of 1.15 GW in 2017 on which models are based |
| **wm** | multiple of wind generation of 6.05 GW in 2017 on which models are based |

---

[1] Email: tonystephensgigg@gmail.com
[2] Department of Engineering and Technology Management, University of Pretoria, South Africa.
   Email: david.walwyn@up.ac.za

1. Introduction

Calculating the contributions of wind and solar generations to the decarbonisation of an electricity system is a simple matter when their combined outputs can be accommodated by the electricity system they serve (Staffell, 2017). Every gigawatt (GW) of combined generation displaces a GW of dispatchable generation, which is likely to be provided by either coal or gas-fired generation (dispatchable generation is defined as that generation required by the electricity system when intermittent sources fail to fully satisfy the difference between demand and base load). It is much more challenging to make such calculations once their combined outputs start to exceed the needs of the electricity system, as has been increasingly the case in the United Kingdom (UK) since 2017 (Gosden, 2021; Gosden, 2017). Excess generation, defined as the surplus of wind and solar generation the electricity system is unable to accommodate, must be curtailed, resulting in a decrease of system efficiency (Stephens and Walwyn, 2018a). Predicting the curtailment of potential energy generation has become an important consideration for energy system design and management.

Unfortunately, steady state models based on annual average performance of the wind and solar fleets are no longer appropriate to this task, since they include, by definition, no information about the variable nature of wind and generations which leads to occasions when their combined generations exceed the needs of the electricity system. They are therefore unable to take account of such occasions, or other occasions when their outputs are close to zero. To accommodate such intermittency, it is necessary to model the wind/solar generation dynamically.

This paper discusses how such modelling may be done for the UK using a dataset of actual 2017 records, taken at 5-minute intervals, from 2017 (Gridwatch, 2021) to construct a Compound Model which retains the real time dynamic interactions between the electricity system and wind/solar generation. This enables annual averages to be calculated, including the calculation of an efficiency measure, the marginal decarbonisation efficiency (MDE). The latter is proposed as an important design constraint, which is likely to determine the upper economic level of investment in the wind fleet.

Finally, we describe how the Compound Model is used to explore the fundamental mathematical relationships between the electrical system and wind/solar generations, resulting in the empirical formulation of a Generic Model which is ideally formulated to be used for planning purposes.

2. Modelling Approach

Although UK wind generation appears to be entirely random, the authors have previously published their finding that, after taking account of variations in annual windiness, annual wind generation histograms for the years 2013 to 2016 showed little statistical variation from year to year. Appropriately scaled wind records for the different years, resulted in very similar predictions of the reductions in efficiency of the wind fleets as they increased in size (Stephens and Walwyn, 2018b). This empirical observation provided the justification for basing UK predictive models on appropriately scaled historic records for a single year (Stephens and Walwyn, 2020a).

Since it is necessary to include solar generation in models, and solar generation records only became available in 2017, it is not possible to base models on pre-2017 records. Another important consideration when deciding on which year's records to base the models is that curtailed generation is not recorded, so it is not justifiable to base models on historic records which include a substantial amount of curtailment. Although a small amount of curtailment was first recorded in 2017 (Gosden, 2017), this would have led to little modelling error when scaled, but curtailment increased substantially in subsequent years (Gosden, 2021). It was decided therefore, to base future model predictions on appropriately scaled 2017 records.

Data for generation and demand on the UK grid is recorded every 5 minutes, resulting in 104,832 records over the period of a year (Gridwatch, 2021). In order to ease the problem of handling such a large data set, it was decided to download the 2017 grid records a week at a time (i.e., 2016 data sets for each week downloaded), creating 52 weekly dynamic models. An advantage of this approach is that a small number of grid records are erroneous but hidden in the large amount of data. When the weekly models are graphed, the errors are easily identified, allowing them to be corrected.

Importantly, the 52 weekly models provide insights into the dynamic nature of wind generation for the wide variety of meteorological conditions experienced during a year. Clearly, there is an enormous difference in the impact of the wind fleet on the electricity system during extended wind lulls, such as a 10-day wind lull in January 2017, when a stationary high pressure system resulted in practically no wind generation in the UK and in most of Europe (Stephens and Walwyn, 2017), as shown in Figure 1, and gales later in the year when wind generation massively exceeded the demand on the electricity system.

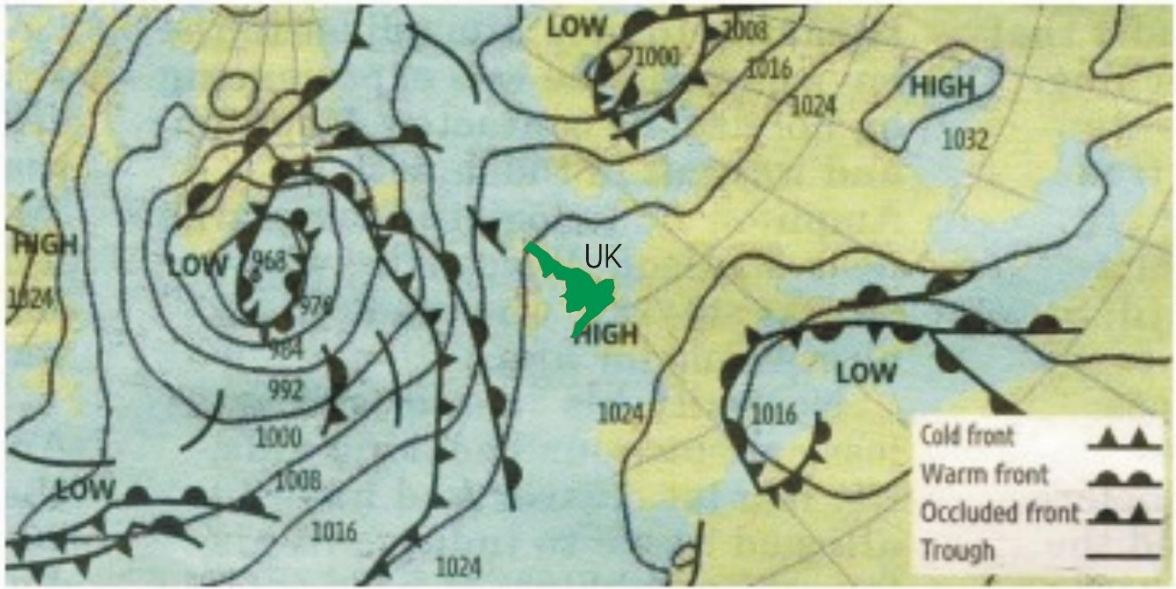

**Figure 1. Weather synoptic for 27 January 2017, during a 10-day period when a stationary high pressure system over the UK (highlighted in green) and Europe, resulted in little wind generation in all the countries affected by the high pressure system**

Currently grid demand varies by typically ± 10 GW during the day. However, we have assumed that it will soon be possible to incentivise a sufficient number of the owners of battery electric vehicles, which will have Vehicle-to-Grid (V2G) capability, to flatten daily demand by charging their vehicles at night and returning some of the stored energy during the day (Needell, Wei and Trancik, 2023; Oldenbroek, Smink, Salet and van Wijk, 2020; Stephens and Walwyn, 2020b). In our view, therefore, it is justifiable to ignore this variation in the medium to long term, for which our models are designed. Additionally, we have previously shown that it is not necessary to model the seasonal variation in demand, since models which assumed annual average demand throughout the year gave similar predictions to those which tracked the seasonal variation (Stephens and Walwyn, 2018b). These two assumptions enormously simplify the modelling, without invalidating its predictions.

Finally, in this section on the modelling approach, we need to explain the concept of Headroom (Hdrm) and how it is applied in the models. Hdrm is essentially the gap between demand and base load, where the latter, in the case of the UK, is provided mainly by nuclear generation. Since nuclear is likely in future to have, as at present, a more privileged access to the grid than wind and solar generation, only Hdrm will be available for the wind and solar fleets to satisfy. This parameter is an important design input value, and fundamental to the understanding of how the models can be applied to energy systems more widely.

Although the aim of the project described in this paper was to develop models capable of assessing the behaviour of the UK electricity system for the wide range of scenarios anticipated in future decades, it is helpful to begin the explanation thereof with an initial application of the models to a single scenario. Such an application is now described.

3.  **Modelling Individual Electricity System Scenarios**

The only published relatively near-term scenario for the UK seems to be that for the year 2035, proposed by the National Grid and published as Future Energy Strategy 2022 (FES 2022) (National Grid Electricity System Operator, 2022). The significance of the year 2035 is that, under the Climate Change Act of 2008 , the government is obliged to take steps to ensure greenhouse gas reductions targets are met by certain dates. The UK's sixth carbon budget requires greenhouse gas emissions to be 78% below their 1990 level by 2035 (Committee on Climate Change, 2020).

Although FES 2022 was a steady state scenario, couched in terms of annual average demand and average sources of generation, it is possible to create dynamic models of the behaviour of the electricity system by assuming a Hdrm value which is consistent with the 2035 scenario, but replacing annual average wind/solar generations with appropriately scaled historic real time records. It is anticipated in FES 2022 that electrical demand in 2035 will be 54.0 GW and nuclear generation 5.48 GW, leaving a Hdrm of 48.5 GW available for the wind and solar fleets to satisfy. The available wind/solar generation proposed for the 2035 scenario were respectively 8.96 and 6.1 times their values in 2035.

### 3.1   Dynamic Modelling of a 2035 Scenario

Weekly dynamic models for the 2035 scenario were created assuming Hdrm of 48.5 GW, and the real time wind/solar generation as recorded in the 2017 but scaled by respectively 8.96 and 6.1 for each 5-minute time interval. The weekly dynamic models were programmed to compare, for each 5-minute time interval, the available wind plus solar generation with the available Hdrm of 48.5 GW, to calculate the following:

- available wind plus solar generation
- wind plus solar generation accommodated by the electricity system (i.e. < 48.5 GW)
- wind and solar generation in excess of that accommodated (i.e. > 48.5 GW)
- dispatchable generation needed to satisfy Hdrm when wind plus solar generation was insufficient to do so
- carbon dioxide generated by the dispatchable generation, assuming it would be provided by combined cycle gas turbines which generate 4.87 Mtes per annum per GW.

The National Grid's model assumes that excess generation (wind plus solar generation not accommodated by the electricity system) would average about 20 GW in 2035, and would be exported via interconnectors which it anticipated would then have a capacity of 25 GW. The UK government's preference was for excess generation to be used to generate hydrogen by electrolysis (Department for Business Energy and Industrial Strategy, 2022). However, only by modelling the 2035 scenario dynamically is it possible to assess whether the dynamic nature of excess generation would make it suitable for either of these uses.

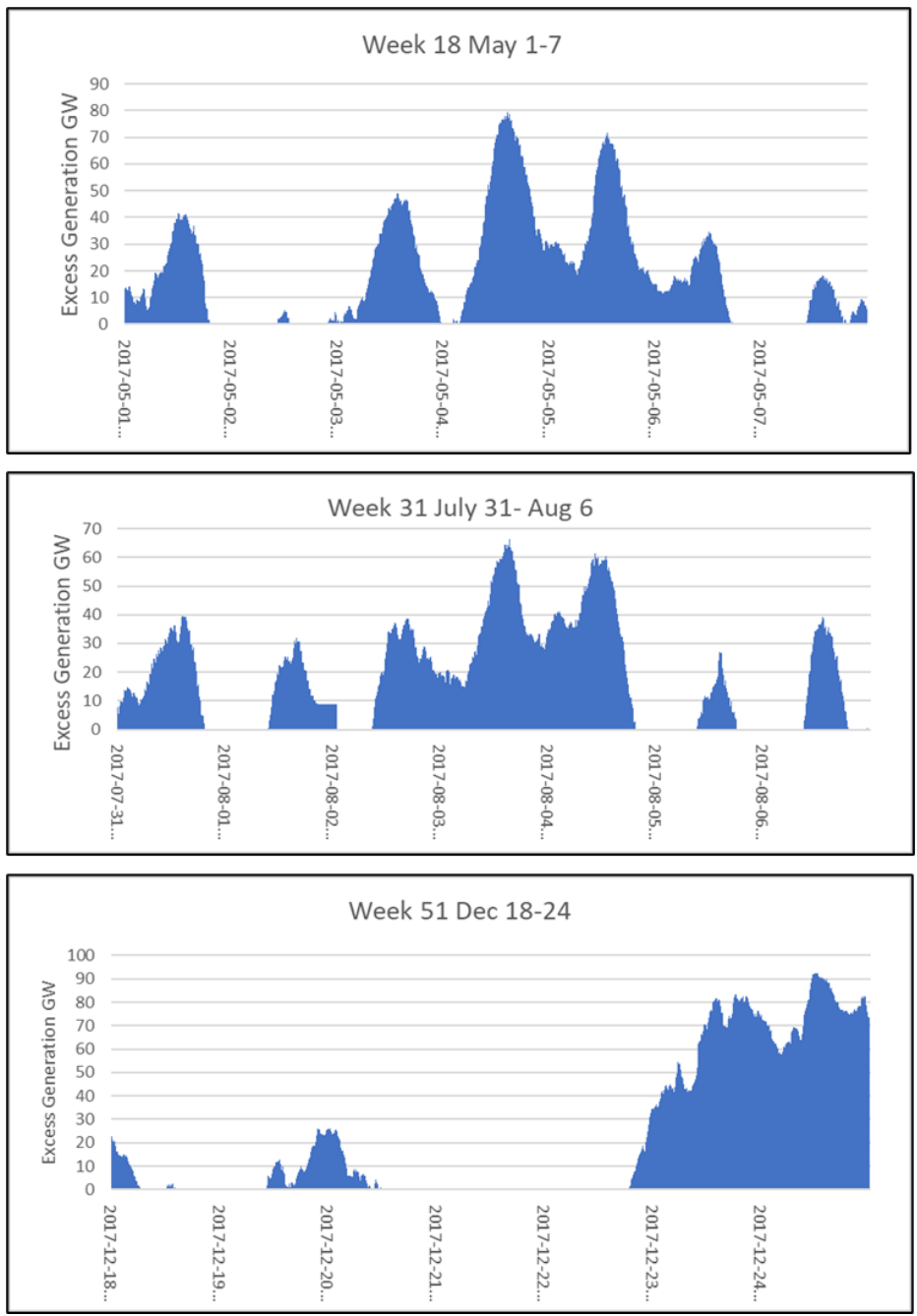

**Figure 2. Predicted excess generation in 2035 based on scaled records of weeks 18, 31 and 51 of 2017**

Figure 2 shows the predicted dynamic nature of excess generation in 2035 based on records of week 18, 31 and 51 of 2017. These weeks were chosen for illustrative purposes, because their average weekly generations were close to the annual average excess generation of 20 GW in FES 2022, with weeks 18 and 31 having significant solar generations each day and week 51 experiencing winter gales on 23rd and 24th December. The 52 weekly models of excess generation revealed that for roughly half the hours in the year we should expect no excess generation for the 2035 scenario, but at other times excess generation to be highly variable and intermittent and up to 100 GW in magnitude.

The variability revealed by the dynamic model predictions of excess generation is what we would expect from a statistical analysis of available wind plus solar generation for the 2035 scenario, shown in the blue histogram of Figure 3. The 2035 scenario requires wind and solar generations to satisfy the Hdrm of 48.5 GW only, but Figure 3 shows some generation of over 150 GW, which implies excess generation of around 100 GW. Similarly, excess generation of just over 90 GW may be seen in week 51 of Figure 2.

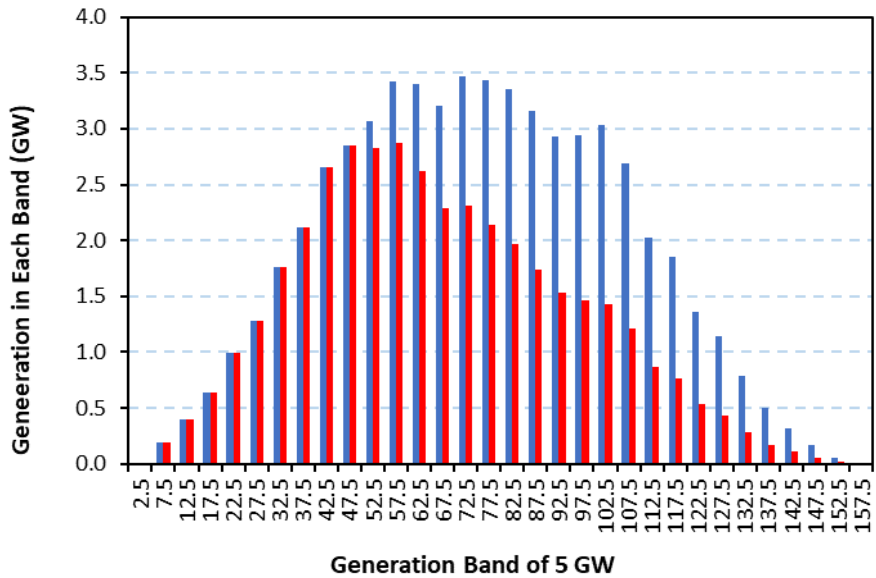

**Figure 3. Histograms of available wind plus solar generation (blue) and wind plus solar generation accommodated by the electricity system (red) for the 2035 scenario.** [3]

The dynamic models suggest that although there might be an average of around 20 GW of excess generation in 2035, its highly dynamic nature means that little of this average 20 GW could be exported using the 25 GW of interconnectors the National Grid proposes for 2035. Another important consideration is whether there is likely to be an export market for UK excess generation which might arise under the 2035 scenario.

This is a question which may be explored by comparing the weekly UK dynamic predictions with the weekly records of the German electricity system in 2017, the latter available on the Fraunhofer web site (Fraunhofer Institute for Solar Energy Systems ISE, 2017). This web site gives graphical historical information of real time German electricity demand, sources of generation, export volumes and prices. The predicted peak UK excess generations in 2035, based on scaled week 18 and week 31 2017 records, which are shown in Figure 2, include sizeable amounts of solar generation. As we might expect, solar generation peaked at the same time in both the UK and Germany. Since Germany was already exporting strongly at such peak times in 2017, it is unlikely that the UK would be able to export at such times in 2035. What reinforces this conclusion is that in 2017 Germany's solar and wind capacities were respectively 42 GW and 55 GW, and it *"plans to expand solar and wind power to 200 GW and 100 GW respectively by 2030, meaning in a few years, Europe's energy market will be swamped with enormous amounts of solar power during mid-days, because Germany will have to sell any electricity that is generated in excess of Germany's own consumption"* (Kastner, 2022).

A 725 km NeuConnect 1.4 GW electrical connection between Wilhelmshaven and Kent is due to come into service in 2028, but is seen by Germany as an opportunity to export some of its enormous surplus to the UK, not an opportunity for the UK to export to Germany (Kastner, 2022). Indeed, records for week 51 of 2017 show that strong gales were simultaneously experienced in the UK and Germany over the period 23rd to 24th December. The German day-ahead auction price was negative for both days, falling to minus 55 Euro per MWh on 24th December. Germany was frequently embarrassed by its excess generation in 2017, sometimes paying neighbours to take it, and this is likely to become even more of a problem by 2030.

---

[3] The way in which these histograms were created will be explained later in the paper.

Further support for our argument that the interconnectors will not assist in redirecting or absorbing excess energy was provided by the energy company EnAppSys, which in July 2023 reported an increase in the number of occasions when UK renewables were causing electricity prices to become negative and it was not possible to export the resulting UK excess generation because European renewables were already themselves generating in excess of demand (Gosden, 2023).

Although the UK government's preference was for excess generation to be used to generate hydrogen by electrolysis (Department for Business Energy and Industrial Strategy, 2022), the predicted dynamic nature of excess generation suggests it would be unsuitable for such use in any major energy intensive process. Even if that were not the case, it must be recognised that producing hydrogen by electrolysis is in its infancy, with only MW sized prototypes currently being developed. A 20 MW electrolyser was sanctioned in 2021 for installation at the Whitelee wind farm in Scotland (ScottishPower, 2021), but the Head of Strategy for BOC electrolyser technology, the company which would operate the Whitelee electrolyser, was not confident about the prospect of success (All Party Parliamentary Group on Hydrogen, 2022).

A 200 MW development electrolyser is to be built near Rotterdam, but it is not planned to be run on excess generation alone - it will have access to the output of the 1,400 MW offshore Hollande Kuste wind farm (Shell, 2022), in addition to solar generation and electrical stored energy. The UK's excess generation will be the result of generation from widely distributed wind farms, and the National Grid has already indicated that it has no intention of installing transmission capacity for handling all the excess generation from the Scottish wind farms which are due to be built in the coming decade (Gosden, 2022). In what follows, it will be assumed that wind plus solar generation which is in excess of the needs of the electricity system will, as has been the case since 2017, continue to be curtailed.

### 3.2 Compound and Histogram Model Predictions of System Relationships

The electricity system's real time relationship with wind/solar generation is that of a low pass filter. It accepts their combined outputs up to the value of Hdrm but is unable to accommodate generation greater than the Hdrm value. It is helpful that there are two different methods of modelling the annual average relationships between future electricity systems and wind and solar generation, and we shall first compare their predictions for the 2035 scenario.

The first modelling approach is to combine the 52 weekly dynamic models discussed in the previous section to create a single Compound Model. For any set of the input variables Hdrm, wm and sm, the model processes the 104,832 annual interactions between the wind/solar generation and the electricity system to calculate their annual average performance.

The other modelling approach is to use the histogram of predicted available wind plus solar generation previously discussed and illustrated as the blue histogram in Figure 3. This was created by:

- scaling the weekly real time historic wind and solar generation records of 2017 by 8.96 and 6.1 respectively
- adding the resulting predicted wind and solar generations for each 5 minute time interval

- creating weekly histograms of available wind plus solar generation in 5GW generation bands for generations ranging from 0 to 165 GW
- averaging the weekly histograms to create the annual ( blue) histogram of available wind plus solar generation of Figure 3.

It is now a simple matter to calculate for each generation band how much of the available wind plus solar generation is accepted by the electricity system. All wind plus solar generation less than Hdrm (48.5 GW for the 2035 scenario) is accepted by the electricity system, but only the fraction of (Hdrm/ centre of generation band) is accepted for bands greater than Hdrm. Thus, for example, the available wind plus solar generation in the generation band centred on 112.5 GW is 2.02 GW, but only 48.5/ 112.5*2.02= 0.87 GW is useful, 1.15 GW being curtailed. The prediction of the available annual wind plus solar generation is the summation of the generations in all generation bands of the blue histogram (61.18 GW), while the useful generation is the summation of generations in the red histogram (40.64 GW). By difference, the excess (curtailed) generation is 20.54 GW.

The predictions of both the Compound and Histogram model for the 2035 scenario are summarised in Table 1, including calculations of carbon dioxide emissions in MT/ annum, assuming dispatchable generation is provided by combined cycle gas turbines which generate 4.87 MT of emissions per GW.

**Table 1. Compound and Histogram Model predictions for the 2035 scenario**

| Parameter | Units | Compound Model | Histogram Model |
|---|---|---|---|
| *Input variables* | | | |
| Hdrm | GW | 48.50 | 48.50 |
| wm | | 8.96 | 8.96 |
| sm | | 6.10 | 6.10 |
| *Model predictions* | | | |
| Available wind plus solar generation | GW | 61.24 | 61.18 |
| GWw+s | GW | 41.06 | 40.64 |
| Excess generation (curtailed) | GW | 20.18 | 20.54 |
| Dispatchable generation needed to satisfy Hdrm | GW | 7.44 | 7.86 |
| *Carbon dioxide predictions* | | | |
| Carbon dioxide generated @ wm | MT/annum | 36.23 | 38.27 |
| Carbon dioxide generated @ wm + 0.1 | MT/annum | 35.69 | N/A |
| MDE | MT/annum/GW | 0.89 | N/A |

The slight differences in the predictions of the two modelling methods may be accounted for by the significantly different sampling frequencies. While the Compound Model calculates excess generation for each of the 104,832 records on which the model is based, the Histogram model carries out this analysis for only 26 generation bands of 5 GW. Although the Histogram Model is useful in providing graphical insights into the variability of wind and solar generations, and in providing confirmation of the predictions of the Compound Model, it is limited in application because histogram productions cannot be automated. The Compound Model on the other hand is ideally suited to carrying out investigatory work, since its 104,832 annual interactions between the electricity system and wind/solar generation are built into its fabric, and the model gives instantaneous predictions of the consequences of varying any of the model variables of Hdrm, wm or sm.

Of importance when planning the appropriate size of future wind fleets is a calculation of the benefit in terms of reduction in carbon dioxide emissions resulting from a further incremental investment in wind generation. We have called this the wind fleet's Marginal Decarbonisation Efficiency, MDE. Although the Compound Model cannot calculate MDE values directly, it may to do so indirectly by running the Compound Model twice for slightly different wm values. Table 1 shows the model being re-run for an increase in wm of 0.1, equivalent to an increase in wind generation of 0.1*6.05 GW. The decrease in carbon dioxide emissions of 36.23 - 35.69 = 0.54 MT/annum results in a predicted MDE of 0.54/0.605 = 0.89 MT/annum/GW.

When the wind fleet was small, each GW of generation would have been accommodated by the electricity system, would have reduced combined cycle gas generation by one GW and hence carbon dioxide emission by 4.87 MT/annum per GW. However, the Compound Model suggests that, if the wind and solar generations are increased to those suggested in the FES 2035 scenario, MDE would be only 0.89 MT/annum per GW (less than 20% of the maximum value). It is beyond the scope of this paper to take a view about what the minimum economically acceptable MDE value might be in future, since it would depend on the comparative efficiencies of other means of reducing the country's carbon dioxide emissions, including improving the thermal efficiency of the UK's poorly insulated housing stock. However, it is hardly credible to suggest the wind fleet should be increased to the point at which it only reduces carbon dioxide emissions by 0.89 MT/annum per additional GW of wind generation.

Although the Compound Model was initially developed to predict the behaviour of the electricity system for individual scenarios, in the next section we shall see how it enabled a structured investigation to be carried out into the fundamental mathematical relationships between the electricity system and its wind and solar generations; an investigation which led to the formulation of a Generic Model.

## 4. Development of a Generic Model

An empirical discovery previously reported by the authors was that the non-dimensional ratio $GWw+s/Hdrm$ could be shown to be a function of MDE but independent of Hdrm for a single sm value (Stephens and Walwyn, 2020a). Although this suggests it might be possible to develop a Generic Model in terms of ratios such as $GWw+s/Hdrm$, the idea was not pursued further because a method of carrying out the necessary analysis was not then available. As will now be shown, it was the development of the Composite Model which provided the means of studying the characteristic relationships between the electricity system and wind and solar generations, and which led to the development of a Generic Model.

### 4.1 Using the Compound Model to Investigate Changes in Wind Fleet Capacity

The first investigation carried out using the Compound Model was of the impact of varying the wind multiple, wm, on wind generation accommodated by the electricity system, GW w, and MDE, for a constant Hdrm of 48.5, assuming no solar generation, i.e. sm=0. To do so, the Compound Model was run 10 times for wm values of 1 to 10, resulting in an array of 10 GW w values, as graphed in Figure 4 left. To an acceptable degree of accuracy, an array of MDE values may be calculated from adjacent members of the GWw wm array using Equation 1, the MDE predictions being shown in Figure 4 right.

Marginal Efficiency at $wm_{n+0.5}$ = (4.87/6.05) * (GWw $wm_{n+1}$ - GWw $wm_n$)     ….. Equation 1

The constant 4.87/6.05 on the right-hand side of Equation 1 is a consequence of each increment of wm representing 6.05 GW of wind generation, and each GW displacing a GW of combined cycle gas turbine generation which would have generated 4.87 MT carbon dioxide/annum/ GW).

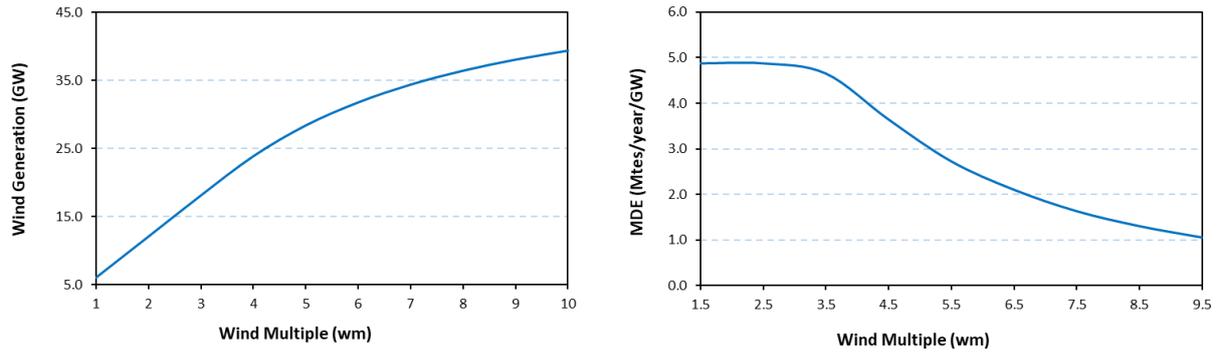

**Figure 4. GWw vs wm (left) and MDE vs wm (right) for Hdrm=48.5 and sm=0**

Figure 4 right reveals a dramatic decrease in effectiveness of the wind fleet in decarbonising the electricity system as the wind fleet increases in size, which is consistent with the prediction of MDE for the 2035 scenario discussed in the previous section. The benefit of using the Compound Model to generate arrays of GW and MDE values for different wm values is that the arrays include the information needed to analyse future electricity systems without the need for further reference to computer models. Although Figure 4 is for a trivial example of only one variable, wm, and sm=0 leading to small arrays of GW w and MDE values , the approach of using the Compound Model to calculate much larger arrays is very useful when more variables are included in the analysis.

### 4.2 Extending the Approach to the Modelling of Additional Variables

The starting point for creating a model capable of predicting the behaviour of the electricity system for the range of scenarios likely to be encountered in coming decades was to create, using the Compound Model, arrays of GWw+s and MDE values for all 150 combinations of Hdrm (30, 40 and 50), wm (1, 2, 3, 4, 5, 6, 7, 8, 9, 10), and sm (0, 2, 4, 6 and 8). By interpolating in the GWw+s and MDE arrays it was possible to generate tables of GWw+s/Hdrm and wm/Hdrm values for integer values of MDE, Hdrm and sm which are reproduced in the table in the Appendix.

The authors have previously published their empirical finding that the ratio GW w+s/Hdrm was sensibly independent of the values of Hdrm for all MDE values for a single sm value (Stephens and Walwyn, 2020a). The tables in the Appendix reveal that both GW w+s/Hdrm and wm/Hdrm are functions of MDE and sm, but sensibly independent of Hdrm viz

$$wm/Hdrm = f_1 (MDE, sm) \quad \ldots\ldots \text{Equation 2}$$
$$GWw+s/Hdrm = f_2 (MDE, sm) \quad \ldots\ldots \text{Equation 3}$$

Since the investigation, which led to derivation of Equations 2 and 3, covered all values of Hdrm, wm and sm values likely to be encountered in future, these equations must have general applicability and therefore are adopted as constituting a Generic Model.

It is possible to derive graphical representations of this Generic Model. While Figure 4 left is a graphical representation of GW w vs wm for sm=0, for a Hdrm of 48.5, Figure 5 is a representation of GW w/Hdrm vs wm/Hdrm for sm=0 but Hdrm values of 30, 40 and 50. The fact that the curves for Hdrms of 30, 40 and 50 lie on top of one another in Figure 5 is simply a consequence of Equations 2 and 3 being independent of Hdrm values, and Figure 5 being a more general representation of Figure 4 left ( i.e. for all Hdrm values).

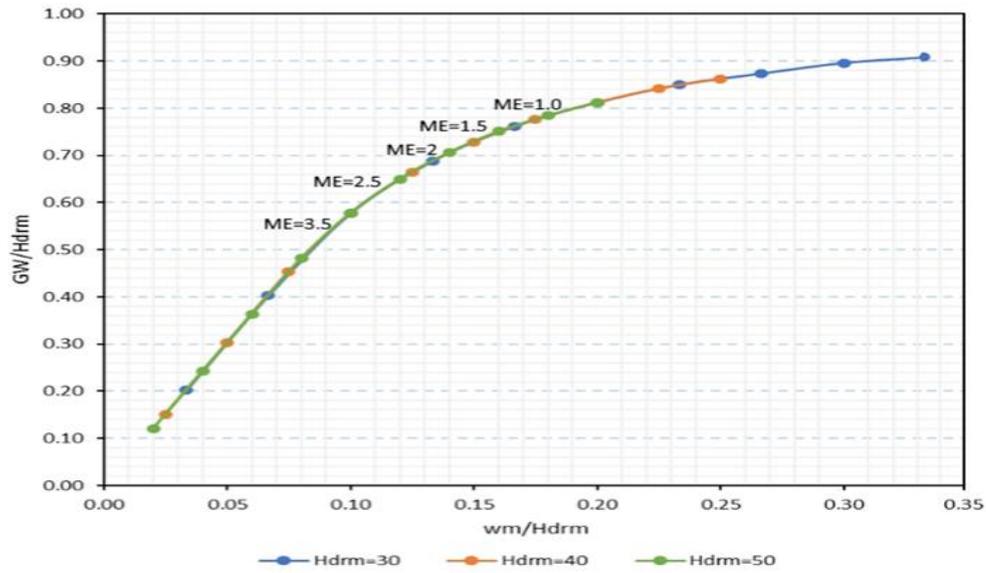

**Figure 5. GW /Hdrm vs wm/Hdrm for Hdrm values of 30,40 and 50**

The analysis is extended further in Figure 6 which shows curves of GW w+s/Hdrm vs wm/Hdrm for sm values of 0, 2, 4, 6 and 8, onto which contours of constant MDE values have been superimposed. Figure 6 therefore represents an alternative graphical representation of the Generic Model of Equations 2 and 3. These equations, in conjunction with the tables of GWw+s/Hdrm and wm/Hdrm in the Appendix, provide an ideal means of investigating individual scenarios, while the graphical representation of the Generic Model in Figure 6 provides useful general insights which are particularly useful for planning purposes.

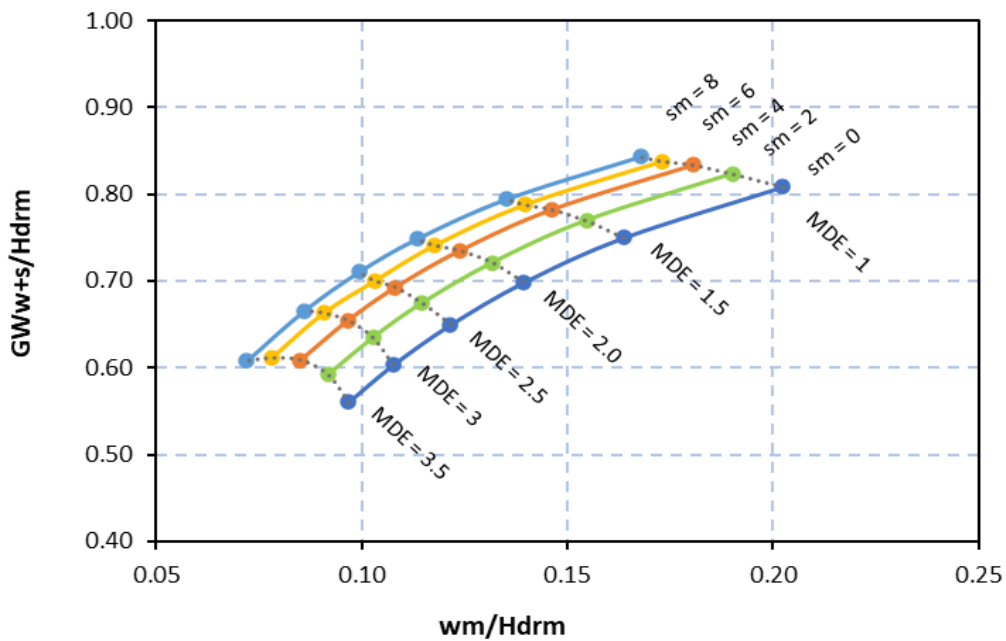

**Figure 6. GW w+s/Hddrm vs wm/Hdrm curves with lines of constant MDE superimposed, which is a graphical representation of the Generic Model of Equations 2 and 3**

## 4.3 Generic Model Predictions

The graphical representation of the Generic Model of Figure 6 provides a means of quantifying the minimum amount of backup dispatchable generation the UK is likely to need, and the consequences for carbon dioxide emissions. The UK government's expectation is for solar generation to be around sm equal to 4 in 2030 and if, for example, the lowest economically acceptable MDE value were 2.5, Figure 6 suggests this would result in GWw+s/Hdrm= 0.7. The dispatchable generation required would be 0.3*Hdrm and would generate 4.87*0.3*Hdrm MT carbon dioxide/annum if provided by combined cycle gas turbine generation. Dispatchable generation of 0.3*Hdrm is an annual average requirement and, to mitigate the occasional lengthy winter wind lull such as that experienced for 10 days in January 2017 (Stephens and Walwyn, 2018a), standby dispatchable generation capacity would be needed equivalent to the full Hdrm value. The annual average utilisation of a standby fleet of combined cycle gas turbines would therefore be 30%.

The conclusion that the UK will need any dispatchable generation provided by gas generation in 2035 runs counter to the steady state models of the National Grid and the Climate Change Committee which show gas generation being discontinued in 2035 in Figure 7. The National Grid's prediction, Figure 7 left, is that there will be around 80 GW of dispatchable generation in 2035, 40 GW provided by electrical storage and 25 GW provided by interconnectors running at full capacity. Unfortunately, neither of these are reliable during winter wind lulls. Only 140 GWh of electrical storage is proposed, so would be exhausted in a little over 4 hours. The authors' previous study of the 10-day wind lull of 2017 caused by the low pressure system of Figure 1 revealed that all major countries were in deficit and that interconnectors could not be relied upon to provide security to the UK energy system. It is difficult to understand how to interpret the Climate Change Committee's ambiguous instruction to the government and to the National Grid for a "*complete end of unabated gas for power generation by 2035, subject to meeting security of supply*".

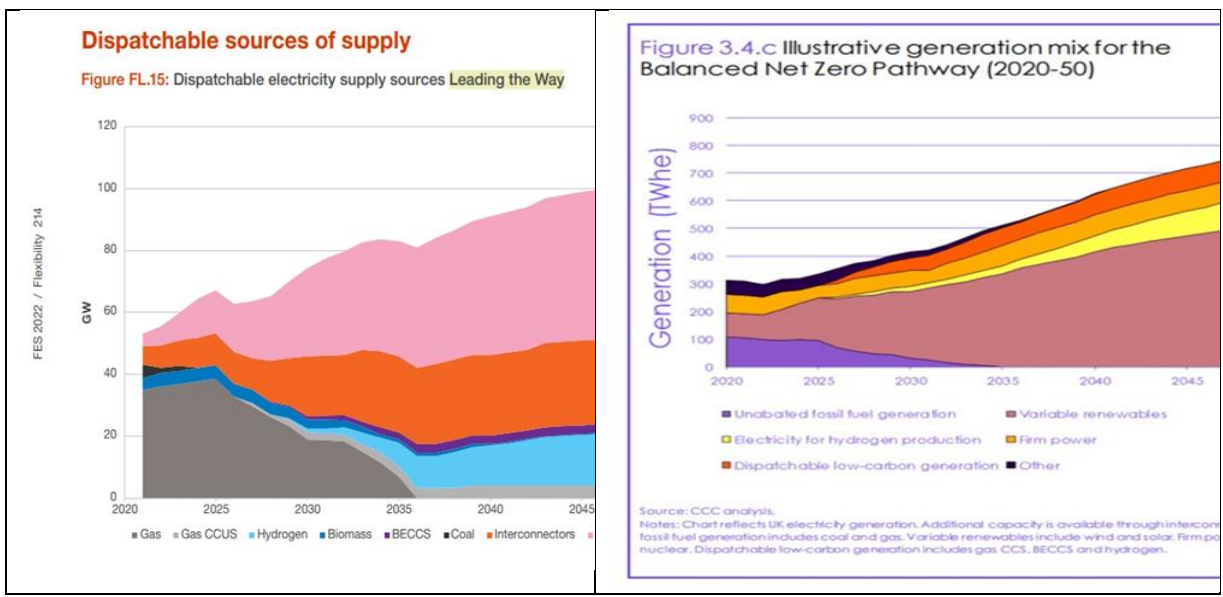

**Figure 7. Suggested sources of dispatchable generation in FES 2022 (left) and in the Climate Change Committee's sixth carbon budget (right)**

The plans of the National Grid and the Climate Change Committee for there to be no back up gas turbine capacity in 2035 is at variant with the plans of other countries equally committed as the UK to transitioning to net zero by 2050. The Fraunhofer website shows that Germany plans to maintain 144 GW of backup gas turbine capacity until 2044 (Fraunhofer Institute for Solar Energy Systems ISE, 2017), and California sees no conflict in using renewables whenever they are available, and backup gas fired generation when they are not available and energy security is threatened (Economist-25 June 2022) .

Another important capability of the Generic Model is its ability to calculate the appropriate sizes of future wind fleets to match the likely future demands on the electricity system. Recent government papers have concentrated on raising wind capacity targets without specifying the electrical demands those capacities are intended to satisfy. Thus, the British Energy Security Strategy paper of 2022 (Department for Business Energy and Industrial Strategy, 2022) proposed that a target of 50 GW offshore capacity by 2030 without specifying what demand the capacity was to satisfy.

With a load factor of 0.45, offshore wind would generate about 22.5 GW which, when added to an anticipated onshore generation of 10 GW, gives a total of 32.5 GW of wind generation, equivalent to a wind multiple, wm, of 32.5/6.05 = 5.27 in 2030. Figure 6 suggests that, for MDE=2.5 and sm=4, wm/Hdrm would be 0.11. The headroom needed to justify this level investment is therefore 5.27/0.11 = 48 GW. Bearing in mind that grid demand fell from 36.5 GW in 2013 to 30 GW in 2022 and that, after considering the reduction in nuclear generation in those years, from 7.53 GW to 5.1 GW, this still gives a reduction in Hdrm from 28.74 GW in 2013 to 24.9 GW in 2022 (Morley, 2023). It seems extremely unlikely that Hdrm will rise from 24.9 GW in 2022 to 48.8 GW in 2030. Should the proposed wind and solar capacity targets for 2030 in the British Energy Security Strategy paper be met (Department for Business Energy and Industrial Strategy, 2022), it seems more likely that, as the Economist warned, the UK would run into a "*wall of stagnant demand*" (The Economist, 2022).

5. **Concluding Remarks**

The intrinsically variable nature of wind and solar generation mean that, as they increase relatively to the needs of the electricity system, an increasing portion of available wind plus solar generation cannot be accommodated by the electricity system. Dynamic models reveal that excess generation will be far too intermittent and variable in magnitude to be put to any beneficial use. As has been the case wind 2017, UK excess generation will continue to be curtailed, with consequent progressive loss in the efficiency with which wind and solar generation are able to decarbonise the electricity system.

A Compound Model is described which is based on appropriately scaled real time records and may be used to calculate a decarbonisation efficiency, the Marginal Decarbonisation Efficiency or MDE. The importance of MDE is that it is likely to determine the upper economic deployment of wind generation. It is shown how the Compound Model may be used to analyse individual electricity system scenarios, such as the National Grid's scenario for 2035, and also to study fundamental mathematical relationships between electricity systems and wind and solar generations which power them.

Such a study resulted in the development of a Generic Model, which should be applicable for electricity system configurations likely to be encountered in future decades and, importantly, may be solved without recourse to computers. The Generic Model is based on the empirical finding that the GW w+s/Hdrm and wm/Hdrm ratios were found to be sensibly independent of Hdrm values. It is not possible to offer a rigorous explanation for this finding, since the ratios are the result of 150 runs on the Compound Model, each of which processed 104,832 appropriately scaled historic wind and solar records. A possible explanation is that, had it been possible to automate the production of wind plus solar histograms, these ratios could have been generated from 150 histograms similar to that shown in Figure 3, rather than 150 Compound Model runs. While real time wind and solar generations appear to be entirely random, Figure 3 shows that a statistical analysis of annual averages result in a Gaussian-shaped histogram. As explained in the text, the annual averages of Table 1 can be generated by treating the histogram as the input to an electricity system which acts as a low pass filter of magnitude Hdrm on each generation band in the histogram.

The main uses of the Generic Model are likely to be in calculating the appropriate amount of wind generation needed to satisfy future UK electricity system demands, and of the dispatchable generation needed to maintain electricity system security when there is insufficient wind and solar generation to do so. The National Grid's steady state models which, until recently have been used as the basis for government policy, give misleading predictions because, by definition, they are unable to analyse the consequences of wind and solar variabilities. Current government policy appears to be to set targets for a significantly increased offshore wind capacity by the year 2030 without consideration of what the demand on the electricity system is then likely to be. As warned by The Economist, there must be a real risk that much of the increased generation will be in excess of the needs of the electricity system (The Economist, 2022). The excess generation will be dynamically unsuitable for putting to any beneficial use and will need to be curtailed. Also, the UK does not appear to have any plans to mitigate lengthy winter wind lulls, such as the 10-day winter wind lull which occurred in January 2017. Germany, which understands the risks of winter wind lulls, or *Dunkelflautes*, plans to maintain a significant amount of gas generation and gas storage into the long term, while the National Grid proposes in FES 2022 to discontinuation of gas generation capability by 2035. The Generic Model provides a means of quantifying how much dispatchable generation will be needed to mitigate the unreliability of wind and solar generation, and suggests that the proposal to have no gas generation capacity by 2035 is an extremely risky one.

*Simulation and Optimization of Wind Farms and Hybrid Systems.* London: IntechOpen, Ch 1, pp 1-19.

Stephens, A. D. & Walwyn, D. R. 2020b. Predicting the Performance of a Future United Kingdom Grid and Wind Fleet When Providing Power to a Fleet of Battery Electric Vehicles. *arXiv preprint arXiv:2101.01065*, pp.

The Economist. 2022. Why Britain is a world leader in offshore wind. *The Economist*, Available: https://www.economist.com/britain/2022/11/24/why-britain-is-a-world-leader-in-offshore-wind [Accessed 29 June 2023].

**Appendix: GW/Hdrm & wm/Hdrm Arrays for Range of Hdrm, MDE and sm Values**

Marginal Efficiency =3.5

| Hdrm | 30 | 40 | 50 | Average | Hdrm | 30 | 40 | 50 | Average |
|---|---|---|---|---|---|---|---|---|---|
| GW/H for sm0 | 0.563 | 0.560 | 0.558 | 0.561 | wm/H for sm0 | 0.097 | 0.097 | 0.096 | 0.097 |
| GW/H for sm2 | 0.602 | 0.591 | 0.583 | 0.592 | wm/H for sm2 | 0.091 | 0.092 | 0.092 | 0.092 |
| GW/H for sm4 | 0.614 | 0.609 | 0.602 | 0.609 | wm/H for sm4 | 0.081 | 0.086 | 0.088 | 0.085 |
| GW/H for sm6 | 0.605 | 0.614 | 0.614 | 0.611 | wm/H for sm6 | 0.072 | 0.079 | 0.083 | 0.078 |
| GW/H for sm8 | 0.592 | 0.613 | 0.618 | 0.608 | wm/H for sm8 | 0.064 | 0.073 | 0.079 | 0.072 |

Marginal Efficiency =3.0

| Hdrm | 30 | 40 | 50 | Average | Hdrm | 30 | 40 | 50 | Average |
|---|---|---|---|---|---|---|---|---|---|
| GW/H for sm0 | 0.607 | 0.603 | 0.601 | 0.604 | wm/H for sm0 | 0.109 | 0.108 | 0.107 | 0.108 |
| GW/H for sm2 | 0.645 | 0.633 | 0.626 | 0.635 | wm/H for sm2 | 0.102 | 0.103 | 0.103 | 0.103 |
| GW/H for sm4 | 0.665 | 0.654 | 0.645 | 0.654 | wm/H for sm4 | 0.094 | 0.097 | 0.099 | 0.097 |
| GW/H for sm6 | 0.669 | 0.663 | 0.656 | 0.663 | wm/H for sm6 | 0.087 | 0.091 | 0.094 | 0.091 |
| GW/H for sm8 | 0.668 | 0.666 | 0.662 | 0.665 | wm/H for sm8 | 0.082 | 0.086 | 0.090 | 0.086 |

Marginal Efficiency=2.5

| Hdrm | 30 | 40 | 50 | Average | Hdrm | 30 | 40 | 50 | Average |
|---|---|---|---|---|---|---|---|---|---|
| GW/H for sm0 | 0.650 | 0.650 | 0.648 | 0.650 | wm/H for sm0 | 0.122 | 0.122 | 0.121 | 0.121 |
| GW/H for sm2 | 0.682 | 0.672 | 0.668 | 0.674 | wm/H for sm2 | 0.113 | 0.115 | 0.116 | 0.115 |
| GW/H for sm4 | 0.704 | 0.691 | 0.682 | 0.693 | wm/H for sm4 | 0.106 | 0.108 | 0.110 | 0.108 |
| GW/H for sm6 | 0.713 | 0.691 | 0.695 | 0.700 | wm/H for sm6 | 0.101 | 0.103 | 0.106 | 0.103 |
| GW/H for sm8 | 0.717 | 0.710 | 0.704 | 0.710 | wm/H for sm8 | 0.097 | 0.099 | 0.102 | 0.099 |

Marginal Efficiency=2.0

| Hdrm | 30 | 40 | 50 | Average | Hdrm | 30 | 40 | 50 | Average |
|---|---|---|---|---|---|---|---|---|---|
| GW/H for sm0 | 0.701 | 0.697 | 0.697 | 0.698 | wm/H for sm0 | 0.140 | 0.139 | 0.139 | 0.139 |
| GW/H for sm2 | 0.729 | 0.720 | 0.714 | 0.721 | wm/H for sm2 | 0.131 | 0.132 | 0.132 | 0.132 |
| GW/H for sm4 | 0.741 | 0.736 | 0.728 | 0.735 | wm/H for sm4 | 0.120 | 0.125 | 0.127 | 0.124 |
| GW/H for sm6 | 0.749 | 0.736 | 0.738 | 0.741 | wm/H for sm6 | 0.114 | 0.118 | 0.121 | 0.118 |
| GW/H for sm8 | 0.755 | 0.745 | 0.744 | 0.748 | wm/H for sm8 | 0.111 | 0.112 | 0.117 | 0.113 |

Marginal Efficiency=1.5

| Hdrm | 30 | 40 | 50 | Average | Hdrm | 30 | 40 | 50 | Average |
|---|---|---|---|---|---|---|---|---|---|
| GW/H for sm0 | 0.753 | 0.748 | 0.749 | 0.750 | wm/H for sm0 | 0.165 | 0.163 | 0.163 | 0.164 |
| GW/H for sm2 | 0.774 | 0.770 | 0.765 | 0.770 | wm/H for sm2 | 0.152 | 0.155 | 0.157 | 0.155 |
| GW/H for sm4 | 0.789 | 0.782 | 0.776 | 0.783 | wm/H for sm4 | 0.143 | 0.147 | 0.149 | 0.146 |
| GW/H for sm6 | 0.796 | 0.782 | 0.785 | 0.788 | wm/H for sm6 | 0.137 | 0.139 | 0.143 | 0.140 |
| GW/H for sm8 | 0.800 | 0.793 | 0.790 | 0.794 | wm/H for sm8 | 0.133 | 0.135 | 0.138 | 0.135 |

Marginal Efficiency=1.0

| Hdrm | 30 | 40 | 50 | Average | Hdrm | 30 | 40 | 50 | Average |
|---|---|---|---|---|---|---|---|---|---|
| GW/H for sm0 | 0.810 | 0.809 | 0.806 | 0.808 | wm/H for sm0 | 0.204 | 0.203 | 0.200 | 0.202 |
| GW/H for sm2 | 0.827 | 0.823 | 0.820 | 0.823 | wm/H for sm2 | 0.187 | 0.191 | 0.192 | 0.190 |
| GW/H for sm4 | 0.839 | 0.833 | 0.829 | 0.834 | wm/H for sm4 | 0.176 | 0.181 | 0.184 | 0.180 |
| GW/H for sm6 | 0.845 | 0.833 | 0.836 | 0.838 | wm/H for sm6 | 0.169 | 0.173 | 0.177 | 0.173 |
| GW/H for sm8 | 0.848 | 0.843 | 0.839 | 0.843 | wm/H for sm8 | 0.165 | 0.168 | 0.171 | 0.168 |